\documentclass[a4paper, twocolumn] {article}

\usepackage{xcolor}
\usepackage{graphicx}
\usepackage[superscript]{cite}
\usepackage[margin=0.585in]{geometry}
\usepackage{textcomp}
\usepackage{amsmath}
\usepackage{comment}
\usepackage{url}
\usepackage{float}
\restylefloat{table}

\usepackage{tikz}

\usepackage{pbox}
\usepackage{array}

\title{Terahertz Oscillator Chips Backside-coupled to Unclad Microphotonics}

\author{Daniel Headland,\footnote{Graduate School of Engineering Science, Osaka University, Osaka 560-8531, Japan}~\footnote{Department of Electronics Technology, Universidad Carlos III de Madrid, Madrid 28911, Spain.}~
Yosuke Nishida,\footnote{ROHM Co.~Ltd., Kyoto 615-8585, Japan.}~
Xiongbin Yu,\footnotemark[1]~
Masayuki Fujita,\footnotemark[1]~
and Tadao Nagatsuma\footnotemark[1]}

\begin{document}

\maketitle

\begin{abstract}
	\textbf{Terahertz technology is largely dependent upon planar on-chip antennas 
	that radiate downwards through the substrate, and so 
	an effective means to interface these antennas with integrated waveguides 
	is an attractive prospect. 
	We present a viable methodology for backside coupling 
	between	a terahertz oscillator chip and a broadband all-intrinsic-silicon 
	dielectric waveguide. 
	Our investigation employs resonant tunneling diodes as compact 
	two-terminal electronic terahertz oscillators, and  
	terahertz waves are observed from 270~GHz to 409~GHz. 
	The fact that this power is accessed 
	via a curved length of silicon waveguide validates successful backside coupling.}
\end{abstract}

\section{Introduction}

Substrateless, all-intrinsic-silicon passive micro-scale photonics (henceforth ``all-silicon microphotonics'') 
 is receiving rapidly increasing interest as a general-purpose
terahertz platform 
\cite{yee2009high, gao2019effective,headland2019bragg,headland2020unclad}.
Devices of this sort are intrinsically efficient because of the near-negligible absorption of high-resistivity 
intrinsic silicon  in the terahertz range,\cite{dai2004terahertz}
in contrast to metals, for which loss increases with frequency  \cite{drude1902elektronentheorie}. 
The absence of a supporting material necessitates patterned cladding for physical support. 
Photonic crystal originally served this purpose,\cite{yee2009high,tsuruda2015extremely}
and subsequent demonstrations employed effective medium,\cite{gao2019effective,headland2020unclad} 
as it offers greater bandwidth.
Related to that point, the omission of any supporting dielectric is also beneficial to bandwidth as
low-frequency leakage into this material is wholly avoided. 

Many useful passive devices have been demonstrated using broadband all-silicon microphotonics, 
including filters \cite{gao2021effective}, 
frequency-division multiplexers \cite{headland2021gratingless}, and sensors enabled by 
integrated optics \cite{headland2021dielectric}.
Lens-enhanced antennas are also readily incorporated to provide an interface 
to free space \cite{withayachumnankul2018all, headland2018terahertz}.
Deep reactive ion etching (DRIE) is employed for fabrication, facilitating
monolithic co-integration of several functional components in a mask-based etching process 
that is amenable to large-volume manufacture. 

Despite the many attractive benefits of the all-intrinsic-silicon platform, there
is one critical limitation:~the direct integration of active devices is simply not possible, and
for this reason, external sources are required.
Initially, terahertz waves were supplied by focusing a free-space terahertz beam 
onto the edge of the silicon wafer \cite{yee2009high}.
This technique is cumbersome and inefficient, but was necessary at the time, 
as the prevailing terahertz measurement apparatus necessarily interfaced with free-space. 
More recently, however, split-block hollow metallic waveguide-based systems for terahertz-range 
measurements have markedly grown in popularity. \cite{hesler2010thz} 
This facilitates measurements that eschew free-space beams altogether by coupling directly to guided waves. 
In the case of all-silicon microphotonics, this connection is achieved by tapering the termination of all-silicon 
waveguides to a point that is then inserted into the hollow metallic waveguide, thereby achieving 
broadband index matching and efficient transfer of terahertz power
\cite{malekabadi2014high,tsuruda2015extremely,hanham2017led, gao2019effective,headland2019bragg,headland2020unclad}. 
Although this is certainly preferable to a focused beam, there are several drawbacks that prevent 
real-world deployment. 
Firstly, split-block packaging increases overall size. 
Secondly, the use of metallic waveguides causes non-negligible conduction 
loss,\cite{drude1902elektronentheorie} which is contrary to one of the
key motivations for the development of the all-silicon microphotonics platform. 
Finally, the interface between the hollow waveguide and the silicon waveguide is sensitive to micro-scale 
misalignment. 
This necessitates the use of micrometer-actuated stages for precise positioning, 
which is clearly not feasible outside of a controlled laboratory environment. 
For these reasons, this approach can be viewed as a useful experimental methodology, but is unsuited to 
practical applications.

It is noted that, at their core, the aforementioned split-block systems
typically house terahertz-range diode devices fabricated on III-V semiconductor 
substrates.\cite{marazita2000integrated,crowe2005opening,hesler2010thz,bulcha2016design}
One of the key challenges faced involves interfacing between the fast-wave mode of the hollow waveguide
and the high-index semiconductor substrate. This requires advanced fabrication techniques such as 
substrate removal,\cite{siegel19992} and transfer to lower-index materials.\cite{marazita2000integrated}
It would therefore be beneficial to connect the terahertz diode directly with micro-scale 
all-silicon passive photonics, from a thick high-index III-V substrate to a high-index silicon waveguide,
and thereby avoid the hollow-metallic-waveguide interface entirely. 
This concept is a form of ``hybrid integration;'' a direct connection between 
heterogeneous integration technologies.

In recent years, considerable effort has been devoted to the concept of terahertz-range hybrid integration,
most notably with resonant tunneling diodes (RTDs).\cite{yu2018integrated,yu2019efficient,yu2020waveguide, yu2021hybrid} 
The choice of RTDs to pilot this development is owed to 
their simplicity, as it is an all-electronic, 
two-terminal mesa that is capable of producing terahertz-range power in response to applied DC bias. 
\cite{suzuki2010fundamental, suzuki2012high, okada2015resonant, maekawa2016oscillation, oshima2017terahertz, kasagi2019large,nishida2019terahertz,iwamatsu2021terahertz}
The absence of complicated oscillator circuitry affords significant design freedom; 
the physical dimensions of the chip substrate can be made very small in order to correspond to the cross-section
of the waveguide, and when patterning the surface, we may concentrate our efforts on the geometry of the coupler. 
And furthermore, RTDs are versatile, as the free-running oscillator 
can be combined with the device's innate nonlinearity in order to yield a highly compact coherent terahertz 
receiver.\cite{nishida2019terahertz,yu2021hybrid}

A planar on-chip metallic coupler is required to make electrical contact with the 
RTD mesa, and to extract terahertz waves therefrom. 
It is reasonable to view this coupler as a kind of antenna, both in terms of its 
physical geometry and in the sense that it interfaces from the chip to the external world, and  
this paradigm allows us to draw upon the extensive, ongoing effort 
that is dedicated to planar on-chip antennas \cite{cheema2013last}. 
The associated literature frequently reports a phenomenon that is of critical importance 
to the present study:~the fields generated by an on-chip antenna are drawn downwards into the 
high-index substrate.
This can cause undesired standing waves in the dielectric cavity that is formed by the 
chip substrate's volume, which negatively impacts bandwidth and front-to-back ratio, 
and produces sidelobes in radiation patterns. 
For this reason, antenna-bearing terahertz-range integrated sources and detectors
are generally backside-coupled directly to a silicon lens at the underside of the 
chip substrate \cite{suzuki2012high, al20121, maekawa2016oscillation, oshima2017terahertz, nishida2019terahertz,kohlaas2020637, harter2019wireless, iwamatsu2021terahertz}.
The close match between the substrate's refractive index and that of the silicon lens 
allows fields to pass naturally through the underside of the chip, thereby suppressing 
standing waves. 

Backside coupling has not previously been applied to hybrid integration with 
silicon microphotonics, owing to challenges in assembly and positioning. 
For this reason, end-fire on-chip antennas have taken 
precedence \cite{yu2018integrated,yu2019efficient,yu2020waveguide, yu2021hybrid}, 
as it is convenient to rest both constituents together upon a single planar surface.  
Although this approach has enabled significant progress, it does pose certain disadvantages. 
The aforementioned phenomenon whereby generated fields are drawn into the substrate can produce
asymmetry in the fields that are output by the antenna, and this translates 
to mode mismatch with the silicon waveguide. 
Another potential issue is that, as the thickness of the substrate determines the relative alignment of the 
antenna and the waveguide, chip substrates must be thinned to a precise bespoke value that is sensitive
to fabrication tolerances.

In this work, we demonstrate a viable technique to backside-couple a terahertz oscillator-bearing 
integrated circuit to an all-silicon microphotonic waveguide.
The oscillator itself is implemented using a micro-scale RTD mesa that is coupled directly to a planar slot
dipole antenna. 
Terahertz power is accessed from the waveguide, and oscillations are observed in the 
from 270~GHz to 409~GHz using a range of different antenna and RTD mesa sizes. 
In this way, we aim to apply the key lessons from previous efforts in the field of planar, 
lens-coupled on-chip 
antennas\cite{suzuki2012high, al20121, maekawa2016oscillation, oshima2017terahertz, nishida2019terahertz,kohlaas2020637, harter2019wireless, iwamatsu2021terahertz}
to a promising emerging passive platform for terahertz waves.\cite{headland2020unclad}

\section{Coupler design}

\begin{figure}[!b]
	\centering
	\includegraphics[width=3.45in]{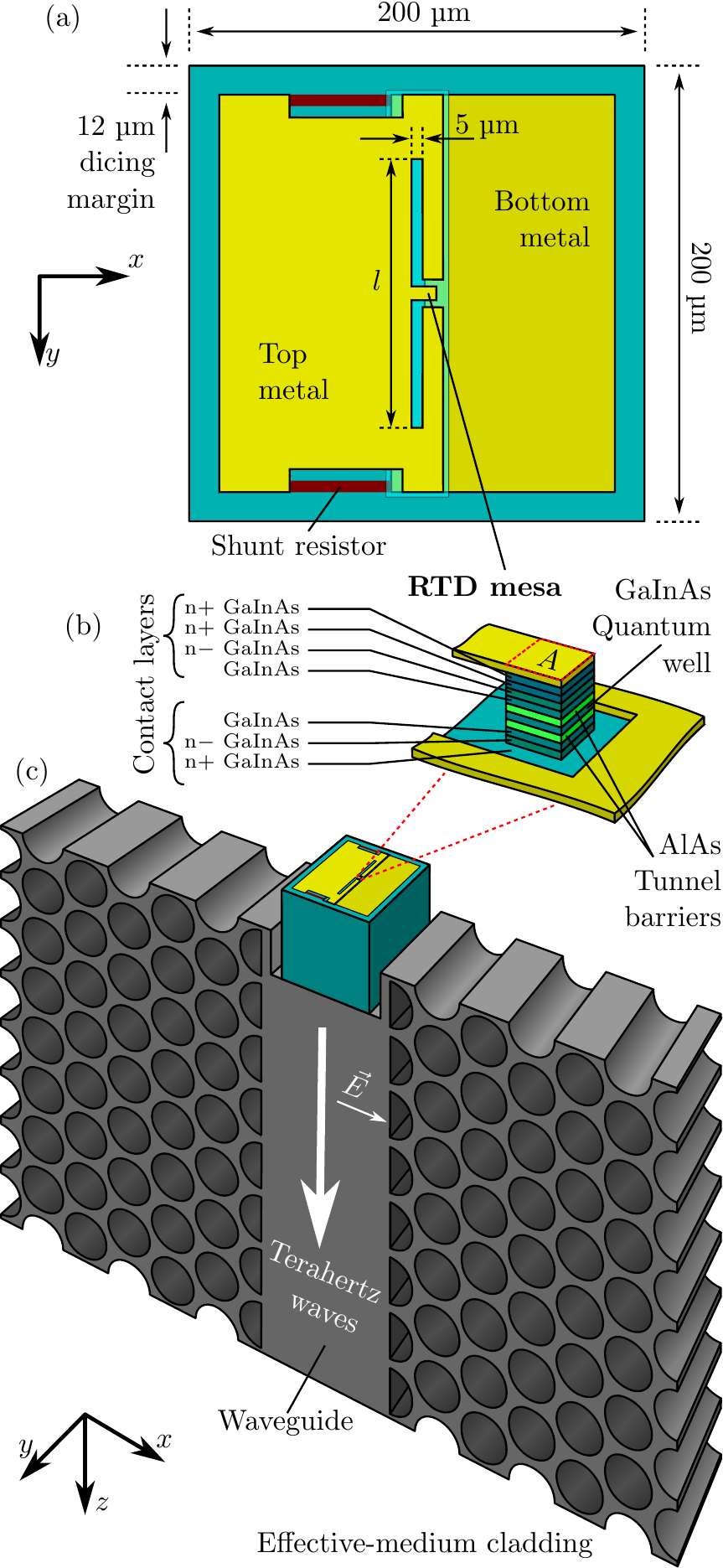}
	\caption{\label{fig:concept} An illustration of the backside-coupling concept, 
	showing (a) the design of the oscillator chip surface, 
	(b) the epitaxial quantum-well structure of diode mesa \cite{diebold2016modeling}, 
	showing mesa area, $A$, and, 
	(c) coupling to a broadband all-intrinsic-silicon terahertz microphotonic waveguide. 
	}
\end{figure}

The main subject of this work is the connection between an active terahertz-range 
chip and a micro-scale dielectric waveguide. 
This section will describe the structure of both components, 
present their physical arrangement, and give numerical analysis of functionality.

\subsection{Oscillator chip}

The RTD-bearing oscillator chip is illustrated in Fig.~\ref{fig:concept}(a). 
The substrate is a $200\times 200 \times 200$~\textmu{}m$^3$ cube of semi-insulating indium phosphide (InP), and 
and the diode mesa is defined on the top surface. 
This epitaxial structure, which is illustrated in Fig.~\ref{fig:concept}(b), consists of 
a gallium indium arsenide (GaInAs)
quantum well that is sandwiched between two aluminum arsenide (AlAs) tunnel barriers. 
The resultant double-barrier tunneling region produces negative differential conductance in the I-V characteristic, 
thereby enabling spontaneous terahertz-range oscillation \cite{diebold2016modeling}.

The RTD is housed within a rectangular slot dipole antenna. 
This particular choice of antenna is well suited to RTDs as it simultaneously serves as resonator and radiator,  
and it is for this reason that the overwhelming majority of noteworthy RTD devices reported in the literature  
employ some form of slot dipole antenna \cite{suzuki2010fundamental, suzuki2012high, okada2015resonant, maekawa2016oscillation, oshima2017terahertz, kasagi2019large}. 
The antenna is also innately compact, as 
slot dipoles tend to be inductive when they are under-sized, i.e.~operated below the frequency at which the
fundamental half-wave mode of resonance occurs, \cite{kominami1985dipole}
which is beneficial to compensate for the capacitance of the diode mesa. 
On the other hand, the closed metallic resonator structure will concentrate 
matter-field interaction with the lossy metal, leading to dissipation. 
This choice of antenna therefore presents a tradeoff pertaining to radiation efficiency, 
compactness, and matching.

The RTD mesa is situated in the center of the slot dipole's length. 
We wish to remark that, although offset-fed slot antennas have previously been proven to offer greater 
output power,\cite{suzuki2012high} here we choose a center-fed configuration in order that the fields that appear 
at the chip's underside will be symmetrical in the $z$-dimension, for greatest correspondence to the 
fundamental mode of an unsupported silicon waveguide. 

In order to allow for electrical biasing, the RTD mesa must make electrical contact with both sides of the slot, 
which must also be insulated from each other. 
Thus, we define two metal layers, namely top- and bottom metal, which are separated by a thin silicon dioxide 
(SiO$_2$) spacer. 
The underside of the RTD mesa makes contact with the bottom layer of metal on the right side of the slot, and a 
narrow SiO$_2$-supported bridge extends from the left side of the slot to make contact 
with the top of the RTD mesa. 
There is potential for loss due to high concentration of current within this bridge, and so its length 
is minimized with a narrow,  5~\textmu{}m-wide slot. 
The slot length, $l$, is left variable in order to tune the resonance frequency. 
Aside from the RTD mesa, the only other conductive pathway between the top- and bottom-metal layers is a pair of
shunt resistors that are intended to suppress undesired parasitic oscillation at lower frequencies.

\subsection{Silicon waveguide}

The all-silicon microphotonic constituent of the design employs 200~\textmu{}m-thick high-resistivity 
float-zone silicon. 
The waveguide track is 300~\textmu{}m-wide, and is clad on both sides with a triangular lattice of cylindrical 
through-holes. 
The pitch of this lattice is 120~\textmu{}m, which is subwavelength over our frequency range of interest, 
and hole radius is 50~\textmu{}m, thereby removing the majority of the silicon to realize a 
low-index effective medium\cite{gao2019effective}.
Terahertz waves are confined within the waveguide core by means of total internal reflection. 
Semi-cylindrical holes are situated on either side of the waveguide in order to reduce longitudinal modulation 
of modal index, as this has the potential to introduce Bragg-mirror effects at high frequencies 
\cite{headland2019bragg}. 

The waveguide is terminated with a 250~\textmu{}m-wide, 100~\textmu{}m-deep rectangular cup
that serves as receptacle for the small RTD-bearing chip, as illustrated in Fig.~\ref{fig:concept}(c). 
There is a 25~\textmu{}m-wide margin on either side of the chip to accommodate fabrication tolerances. 
The base of the cup is oriented towards the direction of propagation in the 
waveguide, i.e.~the positive-$z$ direction, to allow terahertz waves generated by the 
RTD oscillator to pass naturally 
through the substrate and into the silicon waveguide core, as they are both composed of high-index 
semiconductor materials, and are of similar cross-sectional dimensions. 

\subsection{Small-signal simulation}

\label{simulation}

\begin{figure}[!b]
	\centering
	\includegraphics[width=3.45in]{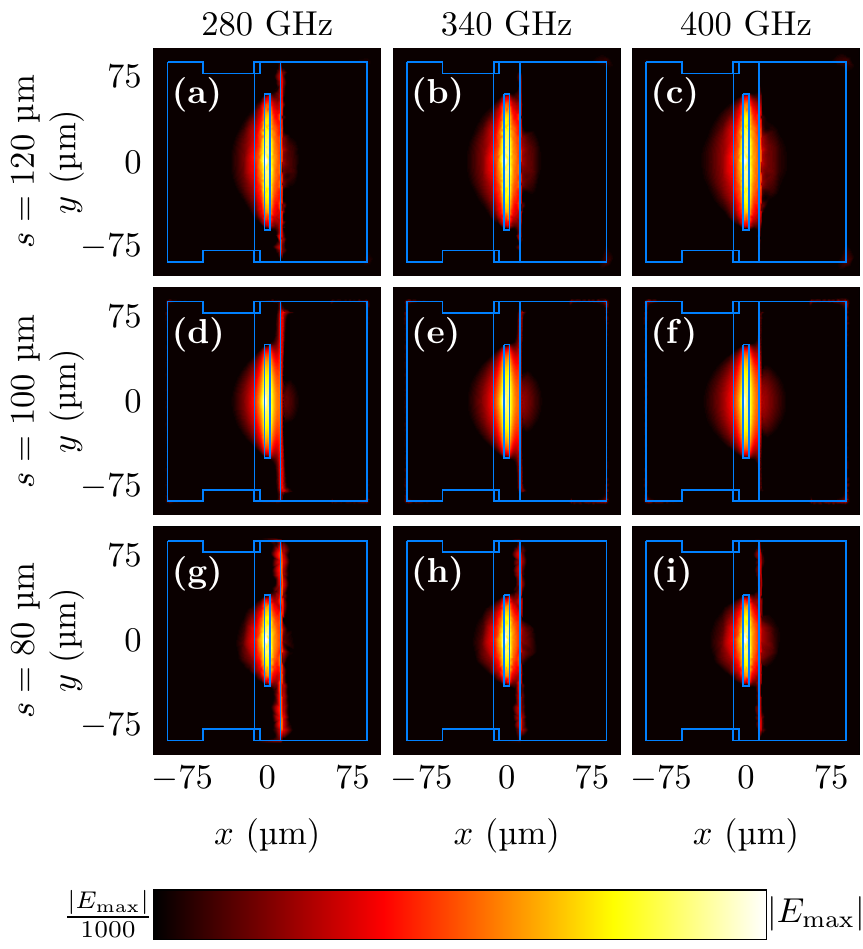}
	\caption{\label{fig:ant-fields} Electric field distribution beneath the chip surface, 
	derived from full-wave simulations of the RTD chip in isolation. 
	Plots are in logarithmic scale, and each is normalized to its own respective maximum.
	}
\end{figure}

\begin{figure}[!t]
	\centering
	\includegraphics[width=3.45in]{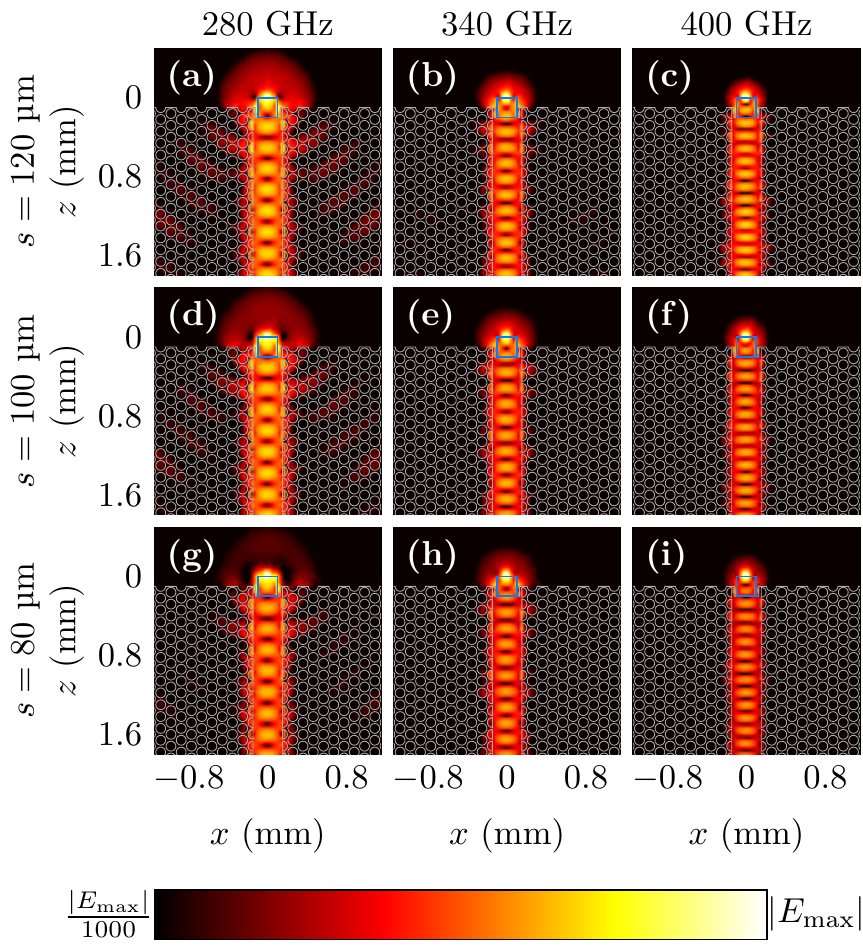}
	\caption{\label{fig:fields} Field pattern results derived from full-wave simulations
	that include both the chip and the silicon waveguide. 	Plots are in logarithmic scale, 
	and each is normalized to its own respective maximum.
	}
\end{figure}

Full-wave simulations using the commercially available CST Studio Suite software package 
are employed to investigate the operation of the RTD chip in isolation.
A 50~$\Omega$ discrete port substitutes for the RTD mesa, and absorbing boundaries 
are affixed to all other surfaces of the substrate in order to prevent nonphysical reflections
from confounding the result.
We simulate a variety of slot sizes, $s$, and the
resultant $E$-field distributions given in Fig.~\ref{fig:ant-fields} show strong 
concentration of electric fields inside the slot dipole in a roughly symmetrical distribution that 
is centered upon the RTD mesa.
It can also be seen that interaction with the shunt resistors is minimal. 
Encouraged by this result, we attach the silicon waveguide to the base of the InP chip, and the
results of this simulation are shown in Fig.~\ref{fig:fields}.
These field patterns qualitatively indicate that a significant 
proportion of the fields injected by the source are indeed commuted 
into the fundamental waveguide mode, and over a broad bandwidth. 
Some radiation loss is also observed, especially at lower frequencies. 
This is to be expected, as the modal fields within dielectric waveguides tend to de-localize into the surrounding
space as frequency is decreased, making low frequencies more liable to scattering loss 
due to irregularities in the surrounding structure.

It is desirable to quantitatively estimate the yield of terahertz power that is conveyed into the waveguide, 
i.e.~the coupling efficiency. 
To this end, a hollow metallic waveguide of inner conductor dimensions 710~\textmu{}m~$\times$~355~\textmu{}m
is attached to the termination of the dielectric waveguide in order to extract terahertz power. 
The silicon waveguide is linearly tapered to a 2~mm-long spike, which is inserted directly into the hollow waveguide 
for index-matching purposes. 
The fundamental, dominant mode of both waveguides closely correspond, and hence the yield of power in the 
hollow metallic waveguide's fundamental mode is a good proxy for the power that is confined within 
the fundamental mode of the dielectric waveguide. 
Note that this interface is not illustrated in this article as it is simply a convenient 
means of interfacing with the simulation model; it is not part of the actual design. 

\begin{figure}[!tb]
	\centering
	\includegraphics[width=3.45in]{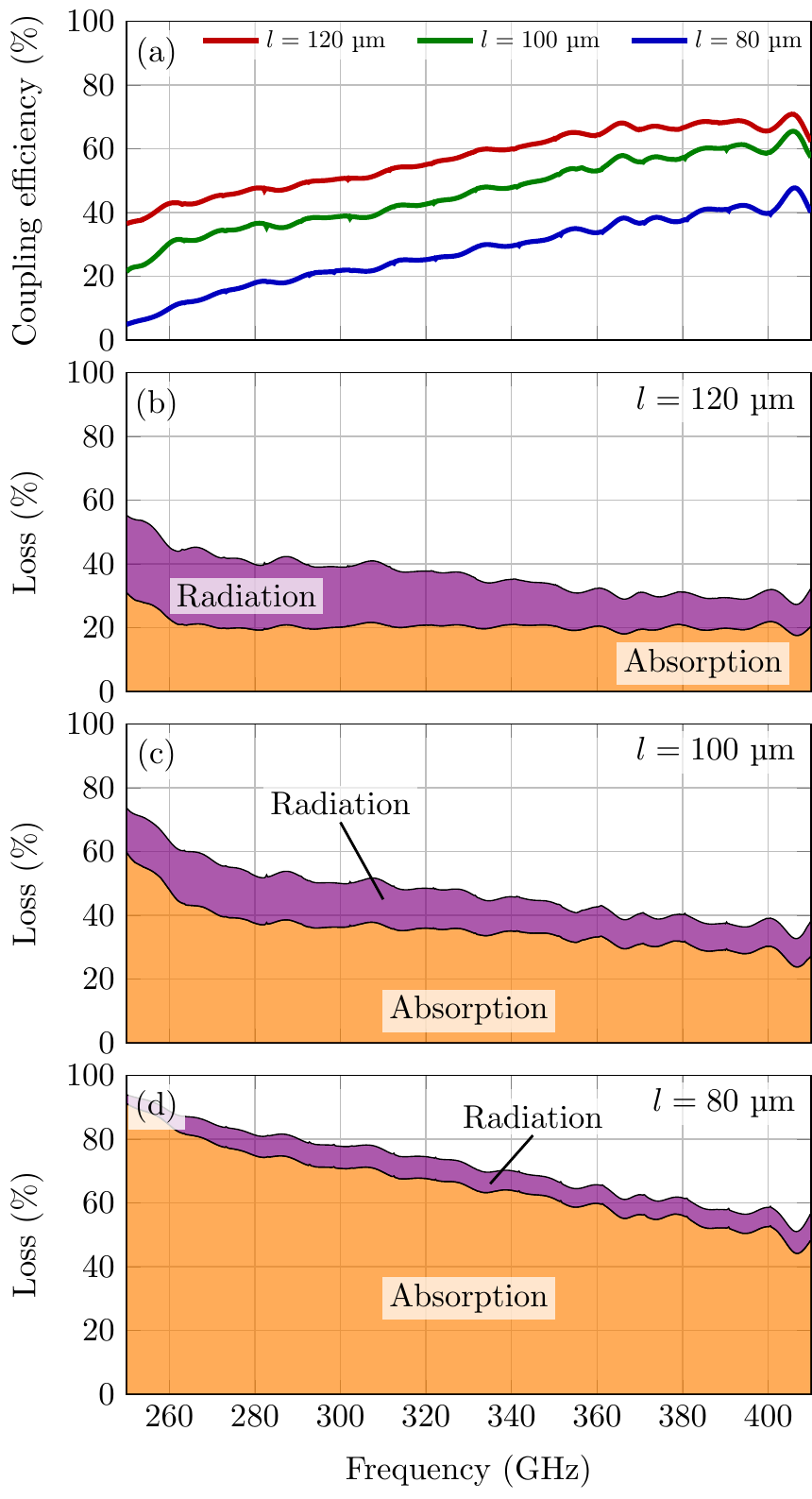}
	\caption{\label{fig:efficiency} (a) Estimate of coupling efficiency from full-wave simulations, 
	assuming conjugate matching between the diode and the slot antenna, 
	and (b)--(d) analysis of contributing 
	factors to loss.
	}
\end{figure}

A two-port $Z$-matrix is extracted from the full-wave simulation, and as this representation does not 
depend upon port impedance, this decouples the RTD mesa from the fictional 50-$\Omega$ source 
that represents it in simulation. 
Thereafter, the $Z$-parameters may be converted into an $S$-parameter 
representation with an arbitrary 
combination of complex impedances at both terminations.\cite{frickey1994conversions}
The hollow-waveguide port is terminated with its own wave impedance in  
order to suppress any reflections therefrom. 
For the discrete port, we employ conjugate matching to the slot antenna, as the cancellation of reactance 
is one of the required conditions for RTDs to oscillate. 
This also represents an ideal case of matching between the diode and the slot, 
 allowing us to concentrate  upon the coupling mechanism between the slot antenna 
and the silicon waveguide, which is the primary focus of this work. 

In order to achieve conjugate matching, we must first evaluate its input impedance
of the slot antenna. 
To this end, the frequency-dependent complex impedance is extracted from the single-port simulations
shown in Fig.~\ref{fig:ant-fields}, is subsequently conjugated, and is used to terminate the 2-port
network. 
The resultant value of $S_{21}$ is normalized against the minor coupling loss at the interface between the 
hollow waveguide and the dielectric waveguide in order to yield the coupling efficiency, and the results are
given in Fig.~\ref{fig:efficiency}(a). 
It can be seen that more efficient transfer of power is achieved in the case of larger slots; peak coupling 
efficiency is 71\% for $s=120$~\textmu{}m, and is 47\% when $s=80$~\textmu{}m.
In the case of all antenna sizes considered, the upper-frequency limit of the 3-dB bandwidth exceeds the 410~GHz 
maximum of the simulated frequency range of interest. 
The lower-frequency limits are $\sim$310~GHz, $\sim$270~GHz, and $\sim$250~GHz for $s=80$~\textmu{}m, 
$s=100$~\textmu{}m, and $s=120$~\textmu{}m, respectively, meaning that relative bandwidth is at least 28\% 
in all cases.

Insight is sought into the causes of loss, and so 
the relative absorption and radiation loss are extracted from the full-wave simulations, 
and re-scaled by the 
total quantity of power that is lost to the network in the $S$-parameter calculation. 
The resultant approximate loss-breakdowns are given in Fig.~\ref{fig:efficiency}(b)--(d). 
It can be seen that radiation and absorption are equivalent for large slots, 
and that absorption loss dominates in the case of smaller slots, especially at lower
frequencies.
This is because, for under-sized slot dipoles, smaller slots will generally have 
greater radiation conductance.\cite{kominami1985dipole}
Under conjugate-matching conditions, this increases the overall current, and hence dissipation, in the metal. 
The high-index substrate exacerbates this phenomenon by further decreasing radiation 
impedance. 
This absorption may also be viewed as the result of concentration of matter-field 
interaction with the lossy metal within an electrically-small closed resonator.

\subsection{Nonlinear device simulation}

The nonlinear RTD component model developed by Diebold et.~al.~\cite{diebold2016modeling} 
is employed to to verify that the proposed chip can generate terahertz waves. 
Complex $Z$-parameters are extracted the full-wave simulations that are presented in 
Section~\ref{simulation}, and are subsequently imported
into harmonic balance simulations using 
the commercially available software package Keysight Advanced Design System. 
This technique iteratively searches for the assumed-sinusoidal solution to the 
nonlinear system that incorporates the RTD and $Z$-parameters in steady-state. 
A selection of antenna sizes is simulated in this way, the 
area of the square RTD mesa, $A$, is also varied,  
and the applied DC voltage, $V_\mathrm{bias}$, is swept for each combination.

\begin{figure}[!tb]
	\centering
	\includegraphics[width=3.45in]{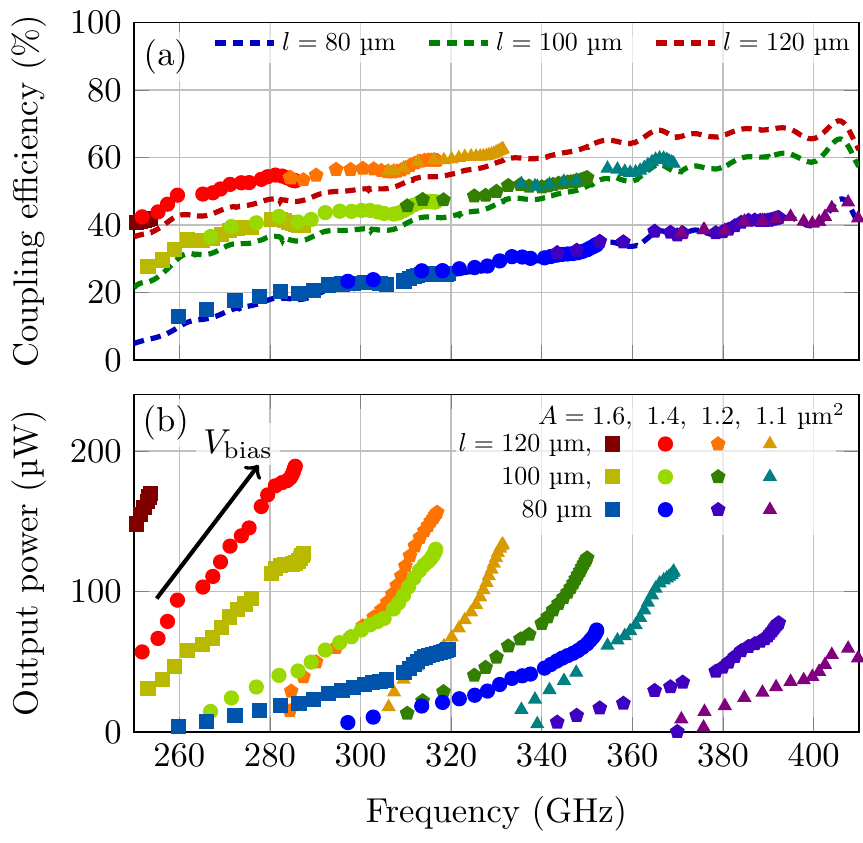}
	\caption{\label{fig:nonlinear} The results of nonlinear circuit simulations, 
	showing (a) coupling 
	efficiency, with the results of full-wave simulations from Fig.~\ref{fig:efficiency} 
	repeated as dashed lines for comparison, and (b) output power. 
	For each case, bias voltage, $V_\mathrm{bias}$ is swept, and the result of 
	all oscillation-producing values is recorded in these plots. 
	}
\end{figure}

We  compute the overall coupling efficiency by comparison of the power generated by the 
RTD with the power delivered to the waveguide port, and the result is shown in 
Fig.~\ref{fig:nonlinear}(a) alongside the conjugate-matching-based
results that are reproduced from Fig.~\ref{fig:efficiency}(a). 
It can be seen that the two methods attain reasonably close agreement, with the nonlinear
simulation yielding slightly higher efficiency. 
This result is somewhat unexpected, as conjugate matching is generally thought
of as the ideal case. 
To explain this result, we observe that conjugate matching delivers the maximum power to 
the lumped impedance of the slot antenna, which incorporates non-negligible losses to 
conduction and backside radiation. 
Thus, the source impedance that maximizes power transfer into the slot may not necessarily
correspond exactly to the ideal case for coupling efficiency. 

The  output power for every combination of mesa area and antenna length
is given in Fig.~\ref{fig:nonlinear}(b). 
It can be seen that, for any given configuration considered,  
both oscillation frequency and output power tend to increase with respect to bias voltage.
In the case of power, we ascribe this to the asymmetrical nature of the I-V curve of the RTD 
device \cite{diebold2017asymmetrical}, and the change in frequency is due to variation in the 
mesa capacitance with respect to applied bias.
Across different design configurations, larger mesas and slot sizes are associated with lower 
oscillation frequencies, as anticipated.
Larger mesas also tend to produce higher output power, as there is a greater area to support 
the flux of current density, and the combination 
of these two facts results in an overall decrease in peak achievable output power 
with respect to  frequency.


\section{Handling}

\begin{figure}[!t]
	\centering
	\includegraphics[]{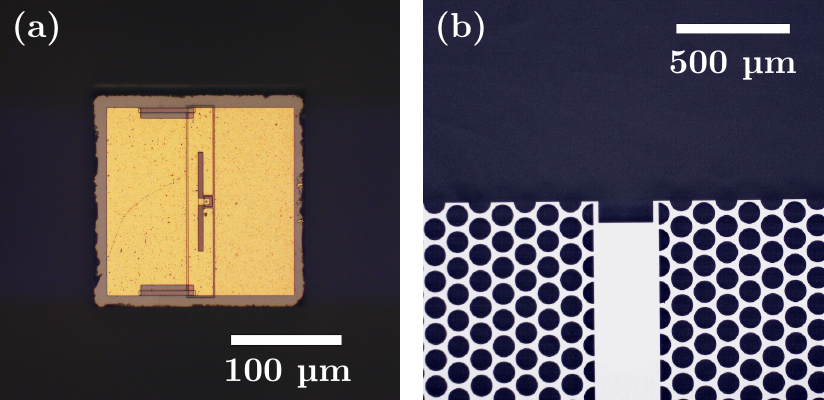}
	\caption{\label{fig:fab} Fabrication, showing (a) a micrograph of a fabricated RTD chip of 
	slot-size $s=93$~\textmu{}m, (b) a micrograph of the cup portion of the silicon waveguide. 
	}
\end{figure}

Fabrication of the heterogeneous integration  technologies is performed at separate 
foundries.
The silicon device employs DRIE, which produces high-aspect-ratio through-holes in 
silicon  using a single mask.\cite{gao2019effective,headland2019bragg,headland2020unclad}
The result is binary; the silicon is either etched-through entirely,
or it is left at full thickness.
On the other hand, the fabrication process of the oscillator chip is more complicated. 
The substrate of the RTD chip is semi-insulating InP upon which the epitaxy shown
in Fig.~\ref{fig:concept}(b) is grown in GaInAs and AlAs. 
The epitaxy is then removed everywhere except for the desired location of the mesa.
A 150~nm-thick bottom-metal layer is formed through a lift-off process.
A 600~nm-thick silicon dioxide layer  is deposited onto the surface of the chip, and is 
selectively removed using reactive ion etching. 
The silicon dioxide insulates the bottom-metal layer from a 800~nm-thick top-metal layer, which is the 
last to be deposited and patterned onto the surface of the chip. 
Finally, the substrate is reduced to its desired thickness, and the individual chips are diced apart using 
a diamond saw. 
Microscope images of both the chip and the silicon component are given in Fig.~\ref{fig:fab}. 

Following fabrication, the two components are subsequently coupled together in the arrangement that is 
shown in Fig.~\ref{fig:concept}(c). 
However, this configuration is not trivial to achieve, as it requires precise placement of fragile, micro-scale 
components---without the option of assembling them together on a single flat surface, as was performed in
previous hybrid integration efforts that employed end-fire antennas 
\cite{yu2018integrated,yu2019efficient,yu2020waveguide, yu2021hybrid}.
We therefore deem it prudent to describe practical issues 
concerning the placement of the oscillator chip in the cup, as well as the convenient 
extraction of terahertz power from the silicon waveguide. 

\subsection{Measurement apparatus}

\label{jig}

\begin{figure}[!b]
	\centering
	\includegraphics[]{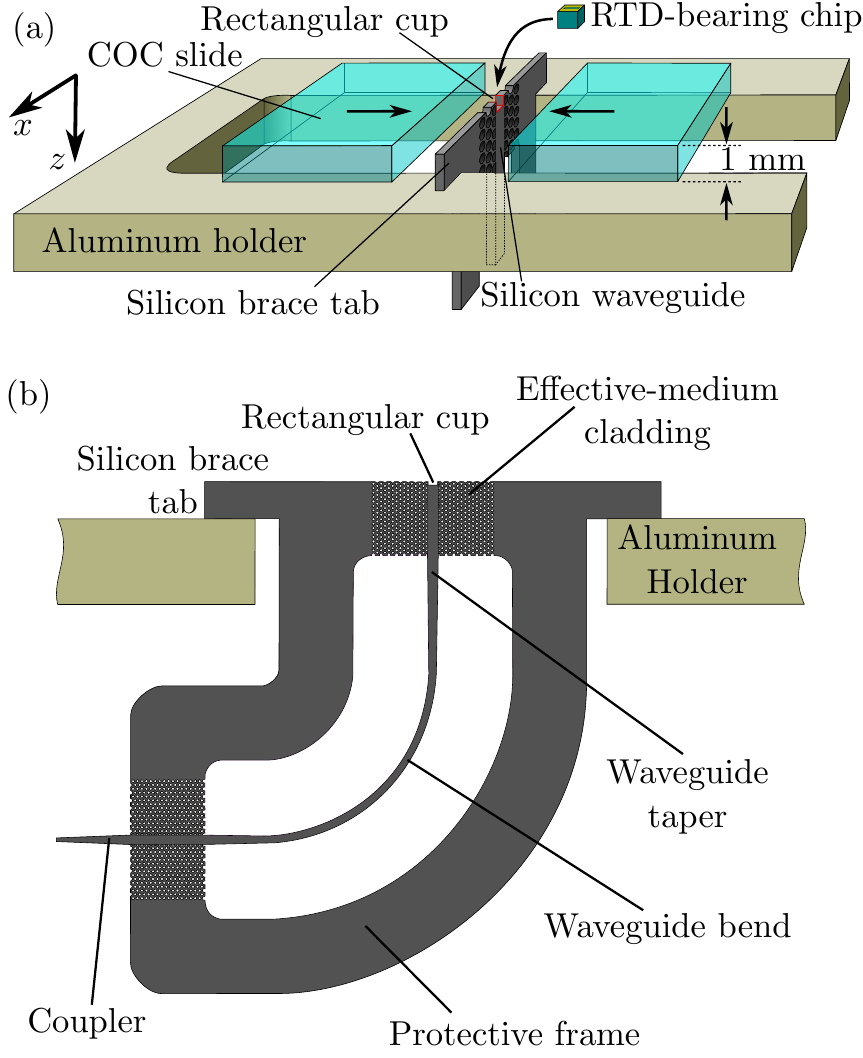}
	\caption{\label{fig:jig} The physical assembly of the measurement apparatus, showing (a) 
	an illustration of the assembly concept (not to scale), and
	(b) a detailed schematic of the silicon portion of the design.
	}
\end{figure}

The surface of the chip must face upwards in order to allow for electrical connection,
with the base of the substrate resting upon the cup.
Thus, the silicon waveguide must be oriented downwards and held firmly in this position, 
and the apparatus illustrated in Fig.~\ref{fig:jig}(a) is devised to serve this purpose. 
A large void is cut into an aluminum holder that is intended to bear the silicon component. 
Following the example of Ref.~\citen{headland2020unclad}, an un-etched frame surrounds 
the silicon waveguide, however in this work, the frame is extended to form a pair of 
rectangular ``brace tabs'' that are level with the cup. 
The undersides of both brace tabs rest upon the flat surface of the aluminum holder, 
on either side of the void,  allowing the waveguide to protrude downwards. 
The silicon cup will therefore face upwards, and be capable of receiving the chip 
in the correct orientation, as intended. 

The oscillator chip must be secured with walls on all four sides in order to prevent it from 
falling out of the cup, and the silicon component is only capable of providing two such walls, 
i.e.~in the positive- and negative-$x$ directions. 
To address this, we deploy a pair of rectangular Cyclic Olefin Copolymer (COC) slides as the remaining 
two walls, and
this particular material is selected as it is both low-loss and low-index in the terahertz range 
\cite{nielsen2009bendable}. 
As shown in Fig.~\ref{fig:jig}(a), the silicon component is sandwiched between the edges of 
the two COC slides, thereby completing the walls of the cup. 
Care is taken to ensure that the thickness of the slides is close to the height of the silicon brace tabs, 
such that equal-height walls enclose the chip on all sides. 
It is also noted that the full-wave simulations presented in Section~\ref{simulation} account for the presence of 
these COC slides. 
The slides are fixed to the aluminum holder with nylon screws. 

After the waveguide passes through the metal holder and protective frame, the effective-medium cladding is 
omitted, leaving a simple silicon microbeam that is both substrateless and 
unsupported.\cite{headland2020unclad} 
This is convenient to allow for bending and routing without disturbing the regularity of the through-hole lattice
that comprises the effective medium. 
The width of the silicon microbeam is 300~\textmu{}m at this point, and modal 
analysis\cite{westerveld2012extension} reveals that undesired higher-order modes are able to propagate 
within this structure for frequencies above $\sim$308~GHz. 
The presence of these modes is clearly undesirable. 
In order to address this, a 3~mm-long linear taper is employed to gradually reduce the waveguide width 
to 200~\textmu{}m.
At the output of the taper, the undesired higher-order modes exhibit cut-off at $\sim$363~GHz, 
and are leaky below $\sim$388~GHz \cite{headland2020unclad}, and hence power that is confined within 
undesired higher-order modes will progressively leak from the waveguide for frequencies below $\sim$388~GHz.
Although this is a form of loss, we deem it preferable to the presence of undesired propagating modes. 

The fact that the silicon waveguide is oriented downwards renders the extraction and detection of terahertz power 
inconvenient in a laboratory setting. 
To rectify this, the waveguide is routed through a 90$^\circ$ circular bend
of radius 4.5~mm. 
The output is directed towards positive-$x$ direction, which is parallel to the 
laboratory table.
Within the frame, the total uncurled length of the waveguide is 11.5~mm.
Finally, the waveguide is passed through the protective frame a second time, 
with effective medium cladding to prevent interaction with the frame, and is 
subsequently tapered to a 2~mm-long spike for progressive index matching as an interface
to the external world. 

\begin{figure}[!t]
	\centering
	\includegraphics[]{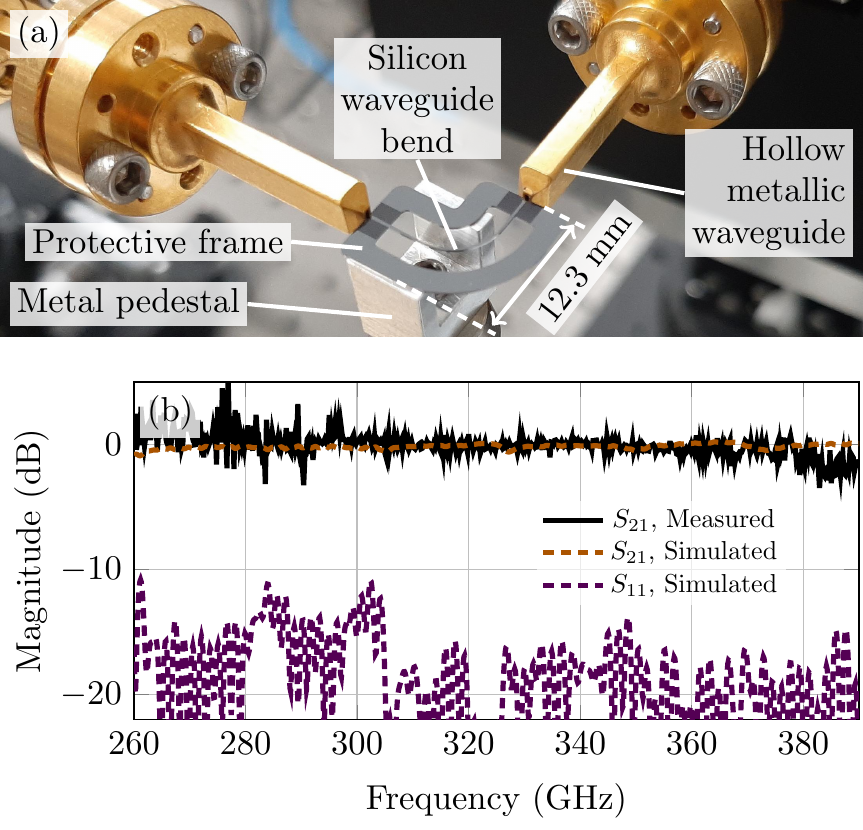}
	\caption{\label{fig:bend}
	(a) Photograph of the silicon bend structure undergoing characterization, and 
	(b) measured estimate of efficiency 
	compared to the results of full-wave simulations. 
	A photograph of the experiment is shown as inset.}
\end{figure}

In order to estimate the losses of the bend structure, a second silicon sample is fabricated
that bears a tapered coupling spike in place of the cup.
As shown in Fig.~\ref{fig:bend}(a), each spike is inserted directly into 
a hollow waveguide for broadband transfer of terahertz power, and the 
sample rests upon a metal pedestal that only makes contact with the protective frame. 
Alignment is achieved using micrometer-driven translation stages guided by the naked eye. 
Terahertz waves are generated with a $\times$9 multiplier that is connected to 
a mm-wave signal generator, and then coupled to the silicon sample. 
Following transit through the bend, the terahertz power is detected using 
an electronic mixer that is coupled together with a $\times$36
multiplier. 
This is connected to the local-oscillator port of a microwave spectrum analyzer for
demodulation and detection of terahertz power. 
Transmission magnitude is normalized by that of an equivalent straight silicon waveguide
 (not shown), and the result is given in Fig.~\ref{fig:bend}(b). 
It can be seen that bending loss is close to 0~dB over the measurable bandwidth.
There is also non-negligible variation in the measured results, of a scale that is 
larger than the bending loss.
We ascribe this to imperfect calibration, as alignment cannot be repeated identically 
between the measurement of the two silicon samples.
This variation is increased for frequencies below $\sim$290~GHz due to innate 
reduction in the dynamic range of the measurement setup. 
The results are compared to full-wave simulations that are implemented with CST Studio
Suite, and the absence of confounding variation renders it possible to observe that 
bending loss is less than 0.9~dB over the measurable frequency range. 
These simulations also indicate that reflection magnitude is below $-$10~dB. 
We therefore conclude that bending loss is not significant. 

\subsection{Positioning the RTD chip}

In our previous work, \cite{yu2018integrated,yu2019efficient,yu2020waveguide, yu2021hybrid} we used 
ordinary tweezers to manually maneuver the chip into a corresponding recess in the silicon device. 
This approach is not possible here due to the small size of the chip. 
In practice, we have found that opening the tweezers does not reliably release the chip, 
likely because the force of the chip's weight cannot overcome the minor electrostatic 
attraction between the chip and the tweezer. 
It must therefore be knocked off by another object, rendering precise placement 
 essentially impossible. 
Furthermore, human hands tend to exhibit minor involuntary tremors---even in the case of 
experienced experimental scientists---and this proved sufficient to damage the silicon 
cup when direct manual placement was attempted.
Thereafter, we placed the chip on the slide surface adjacent to the cup, with the 
aim of nudging it into the cup with the tip of the tweezers. 
This solved the problem of involuntary tremors, but the chip unfortunately 
tended to tilt and rotate as it fell. 
As a consequence, one of the sides of the chip substrate would make contact with the top of one 
of the walls of the cup, leaving it in an improper, diagonal orientation, without 
flush contact between the bottom of the chip and the base of the cup. 
Any attempt to rectify this by applying tweezer-pressure would either result in damage, 
or to the chip itself catapulting away and being lost. 

\begin{figure}[!t]
	\centering
	\includegraphics[]{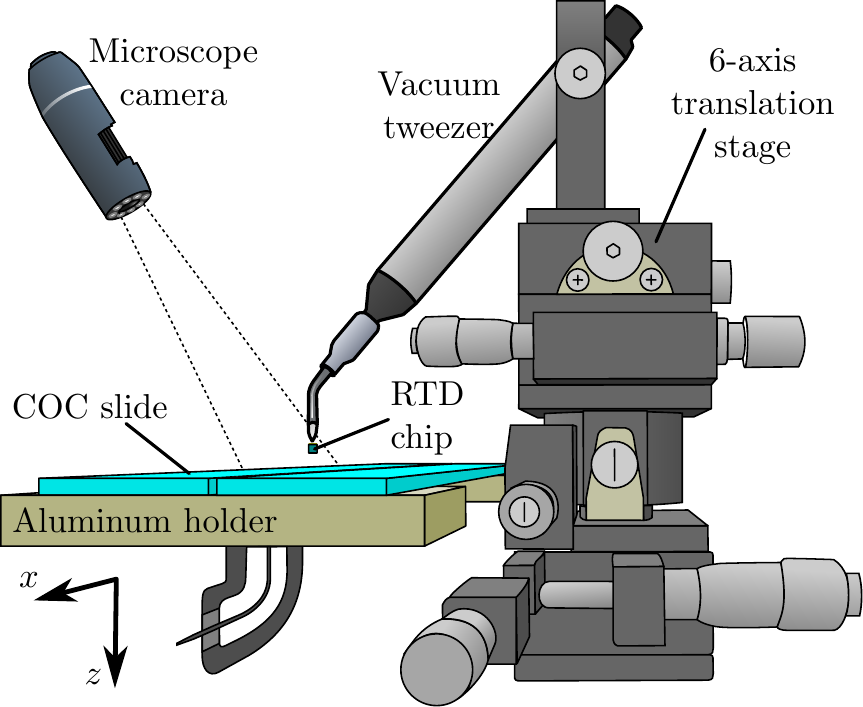}
	\caption{\label{fig:handling} Chip-handling apparatus that serves to place the 
	RTD-bearing chip into the cup that terminates the silicon waveguide. 
	}
\end{figure}

Due to the challenges of ordinary-tweezer placement, we develop an alternative, vacuum-tweezer-based technique 
to place the chip into the cup, as illustrated in Fig.~\ref{fig:handling}. 
A 130~\textmu{}m-diameter polyoxymethylene vacuum-tweezer tip is attached to a micrometer-actuated 6-axis 
manual translation stage that is placed adjacent to the aluminum holder, and a diagonally-oriented microscope camera 
observes the top surface of the COC slides. 
During placement, a given chip is retrieved from its storage case, and placed upon one of the COC slides using
ordinary hand-held tweezers. 
The RTD chip is released from the tweezer by gently brushing it against the slide. 
Thereafter, the same hand-held tweezers are employed to re-orient the chip such that the antenna faces upwards, and 
with the correct rotation relative to the cup. 
Thereafter, the translation stage is employed to bring the tip of the vacuum tweezer into the vicinity of the chip.
Once contact is made with the top surface of the chip, vacuum suction is switched on, and the translation 
stage raises it upwards (i.e.~the negative-$z$ direction). 
The chip is then positioned above the silicon component, and gradually lowered into the cup. 
Finally, the suction of the vacuum tweezer is switched off, and the chip is released in its 
desired position. 

Adequate visibility is vital for the success of the chip-positioning procedure.
The naked eye is unable to perceive the relevant micro-scale features, and hence the microscope camera is an 
essential component of the experimental setup that is illustrated in Fig.~\ref{fig:handling}. 
In our previous work, \cite{yu2018integrated,yu2019efficient,yu2020waveguide, yu2021hybrid} 
the camera was oriented directly downwards, and so the setup was viewed from above. 
Here, however, the vacuum-tweezer tip would block the view of the chip. 
We therefore orient the microscope camera diagonally, but this arrangement 
presents its own challenges, as it projects the $y$- and $z$-dimensions onto each other, thereby 
introducing visual confusion. 
The operator of the experiment must therefore rely on visual cues such as reflections and shadows in order 
to resolve the ambiguity.

\section{Experiment}

\begin{figure}[!b]
	\centering
	\includegraphics[]{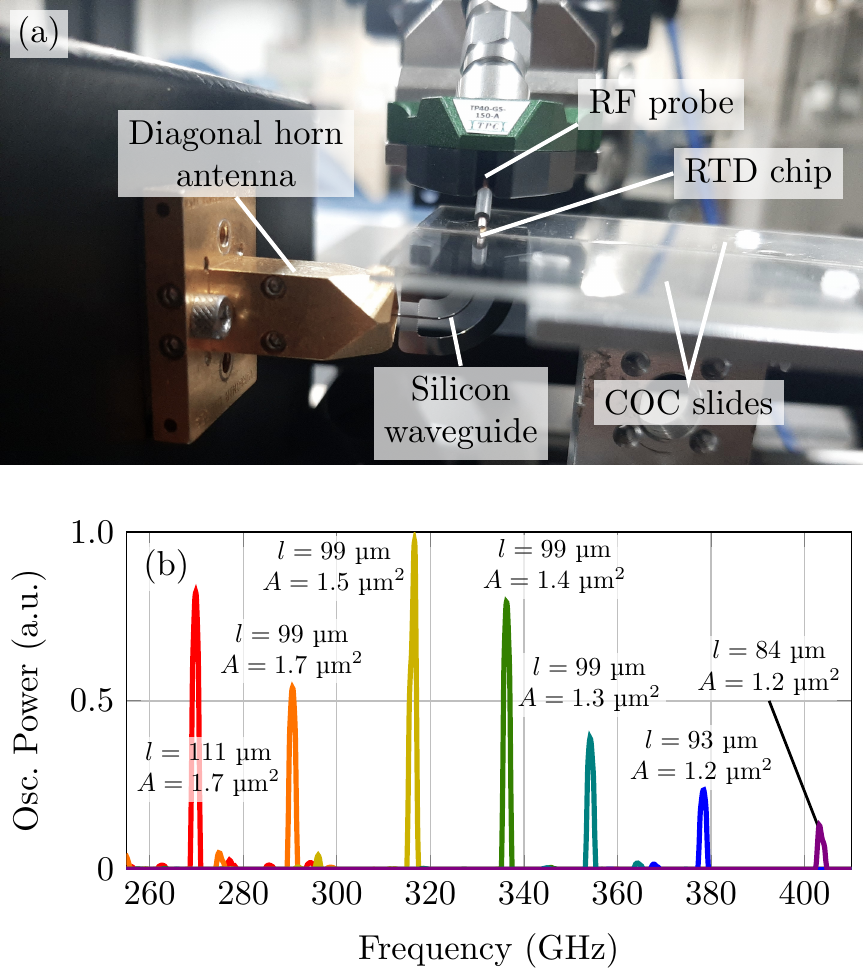}
	\caption{\label{fig:osc} Experimentation, 
	(a) an annotated photograph of the experiment, and 
	(b) measured oscillation from several chips of varying antenna and mesa sizes. 	
	}
\end{figure}

Having positioned the chip into the cup, the vacuum tweezer stage is removed, 
and is replaced with an 150~\textmu{}m-pitch RF GS probe that provides an 
electrical connection. 
The tapered spike coupler at the output of the dielectric waveguide is oriented towards 
a diagonal horn antenna, which is coupled to 
the electronic terahertz-range receiver that is described in Section~\ref{jig}.
A photograph of this experiment is shown in Fig.~\ref{fig:osc}(a). 

In previous related work,\cite{gao2019effective,headland2019bragg,headland2020unclad} 
the tapered spike has been inserted directly into a single-mode hollow metallic 
waveguide, offering significantly more efficient coupling.
However, a horn antenna is preferred here as the silicon component may 
shift slightly during chip positioning, which could lead to breakage if the tapered 
tip were situated within a hollow metallic waveguide.
Thus, the use of the horn antenna facilitates safe and rapid probing of several different 
chips consecutively. 

The RF probe supplies DC voltage to the chip,  causing spontaneous terahertz-range 
oscillations in the RTD, and resultant spectra are observed with the spectrum analyzer. 
A total of 132 chips of several different antenna and mesa sizes were tested in this way, 
and 46 were found to oscillate---a yield of 35\%. 
A selection of acquired spectra were smoothed with a 1-GHz raised cosine window, 
and the result is shown in Fig.~\ref{fig:osc}(b).
It can be seen that terahertz-range oscillations are observed over a 139-GHz span from 270~GHz 
to 409~GHz. 
Due to the usage of a diagonal horn antenna, it is not 
possible to calibrate these measurements, and hence the spectra are presented in 
arbitrary units. 
Nevertheless, these relative-power measurements do offer some insights. 
It can be seen that output power is greater for oscillations below $\sim$350~GHz. 
This may be related to reduced coupling efficiency for smaller antennas, as shown 
in Fig.~\ref{fig:efficiency}(a) and Fig.~\ref{fig:nonlinear}(a), 
as well as with the presence of undesired higher-order leaky modes that begin to propagate 
at $\sim$360~GHz. 
It is also noted that high-frequency oscillation necessitates smaller RTD mesa areas, 
which generate less power, as shown in Fig.~\ref{fig:nonlinear}(b).

\section{Conclusion}

We have experimentally demonstrated backside-coupling between a terahertz oscillator chip 
and a broadband, unclad, unsupported all-intrinsic-silicon microphotonic waveguide.
The chip is placed into a silicon cup at the termination of the silicon waveguide using 
a specialized handling procedure that we have reported in detail.
Terahertz waves between 270~GHz and 409~GHz were extracted from the termination 
of the silicon waveguide, thereby confirming transfer of power from the chip, and
validating this  proof-of-concept for terahertz-range hybrid integration 
using a backside-coupled chip. 

\begin{figure}[!t]
	\centering

	\includegraphics[width=3.45in]{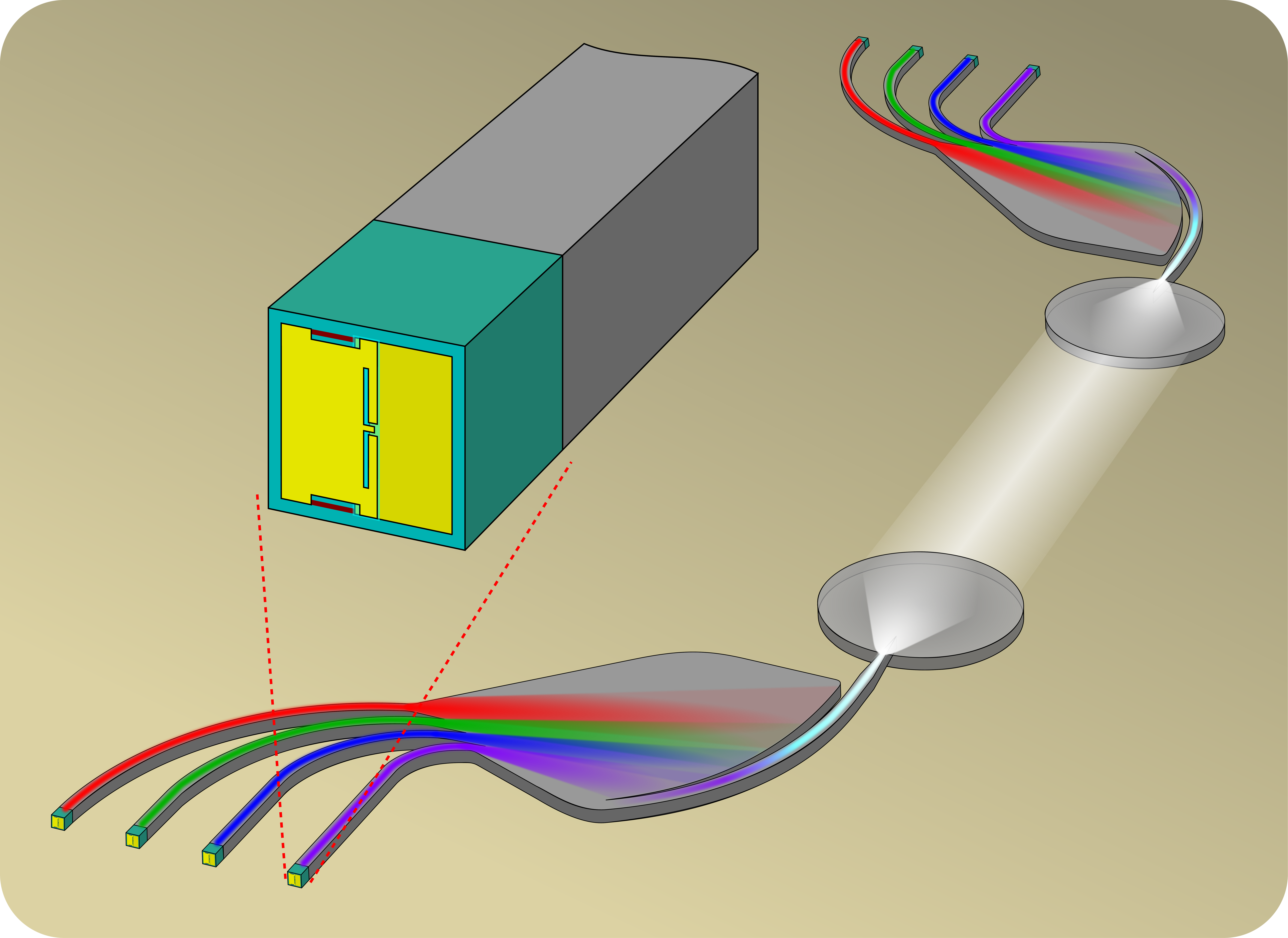}

	\caption{\label{fig:system_future}  
	Concept of a terahertz wireless link 
	using multi-channel microphotonic transceiver systems that 
	combine the hybrid-integration technique that is 
	reported in this article together with a monolithically-integrated 
	wide-aperture antenna \cite{headland2018terahertz}
	and multiplexer \cite{headland2021gratingless}.}
\end{figure}

The passive microphotonic component that is deployed in the context of this work is  a simple 
90$^\circ$ waveguide bend for routing purposes. 
We wish to remark that 
more-sophisticated and useful devices are readily available. 
For instance, frequency-division  multiplexers \cite{headland2020unclad, headland2021gratingless, yu2021hybrid}
may be employed for channelization, and  
all-silicon wide-aperture antennas \cite{withayachumnankul2018all, headland2018terahertz} may also 
offer an interface to free space.
The resultant system would be a multi-channel wireless transciever module for  
terahertz-range high-speed communications, as illustrated in Fig.~\ref{fig:system_future}.

In principle, this backside-coupling technique is applicable to frequencies as high as 
1~THz, as both RTD oscillators\cite{maekawa2016oscillation} and substrateless 
all-silicon microphotonics\cite{yee2009high}
have been demonstrated at higher frequencies than those observed in the present work. 
That said, this will necessitate a reduction in chip size and waveguide cross-section dimensions, 
which may render placement more difficult. 
Further investigation is therefore required in order to determine the practical 
upper-limit of 
high-frequency operation in this manner of backside-coupled arrangement. 

Although our present investigation primarily concerns RTD devices, the assembly
technique is applicable to any given terahertz two-terminal active device, 
including Schottky barrier diodes\cite{marazita2000integrated, bulcha2016design} 
and laser-excited optoelectronic 
devices\cite{kurokawa2018over,harter2019wireless,kohlaas2020637}.
Likewise, there is no special significance to the slot dipole antenna other than its
suitability to RTDs. Instead, we may enhance bandwidth and efficiency by using 
log-periodic, \cite{rivera2015dielectric} 
ring,\cite{al20121} or bow-tie antennas.\cite{harter2019wireless}
As such, this article has reported a pilot study of a general-purpose terahertz
hybrid-integration strategy to combine diverse active source and detector terahertz 
chips with broadband and efficient all-silicon microphotonics.

\section*{Acknowledgment}


This work was in part supported by the Core Research for Evolutional Science and Technology (CREST) 
program of the Japan Science and Technology Agency (JPMJCR1534 and JPMJCR21C4), the commissioned research 
by National Institute of Information and Communications Technology (NICT), Japan (03001), and KAKENHI, 
Japan (20H00249).

The authors wish to thank Mr Yuichiro Yamagami and Mr Shuya Iwamatsu for their preliminary efforts
towards manual chip placement. 

\small

\bibliographystyle{ieeetr}

\end{document}